# Radiation Reaction in High-Intensity Fields

Keita Seto

*Extreme Light Infrastructure – Nuclear Physics (ELI-NP) / Horia Hulubei National Institute for R&D in Physics and Nuclear Engineering (IFIN-HH),*

*30 Reactorului St., Bucharest-Magurele, jud. Ilfov, P.O.B. MG-6, RO-077125, Romania*

* keita.seto@eli-np.ro

. . . . . . . . . . . . . . . . . . . . . . . . . . . . . . . . . . . . . . . . . . . . . . . . . . . . . . . . . . . . . .

After the development of a radiating electron model by P. A. M. Dirac in 1938, many authors have tried to reformulate this model so-called "radiation reaction". Recently, this effect has become important in ultra-intense laser-electron (plasma) interactions. In our recent research, we found the stabilization method of radiation reaction by the QED vacuum fluctuation [PTEP **2014**, 043A01 (2014), PTEP **2015**, 023A01 (2015)]. On the other hand, the modification of the radiated field by highly intense incoming laser fields should be taken into account when the laser intensity is higher than $10^{22}$W/cm$^2$, which could be achieved by the next generation ultra-short pulse 10PW lasers, like the ones under construction for the ELI-NP facility. In this paper, I propose the running charge-mass method for the description of the QED-based synchrotron radiation by high-intensity external fields with the stabilization by the QED vacuum fluctuation as an extension to the model by Dirac.

. . . . . . . . . . . . . . . . . . . . . . . . . . . . . . . . . . . . . . . . . . . . . . . . . . . . . . . . . . . . . .

## 1. Introduction

With the rapid progress of ultra-short pulse laser technology, the maximum intensities of these lasers has reached the order of $10^{22}$W/cm$^2$ [1, 2]. If the laser intensity is higher than this, strong radiation may be generated from a highly energetic electron. Accompanying this, "radiation reaction" as the feedback from radiation to an electron's motion can have a strong influence on electrons in plasmas [3]. One of the facilities which can achieve these regimes, Extreme Light Infrastructure - Nuclear Physics (ELI-NP) will feature two 10PW (approximately $10^{24}$W/cm$^2$ at tightest focus) class lasers [4-6]. At these intensity levels, the radiation reaction must be taken into account in the laser-plasma experiments carried out. The original model of radiation reaction, described by the Lorentz-Abraham-Dirac (LAD) equation [7], has a significant mathematical difficulty which is an exponential divergence $dw/d\tau \propto \exp(\tau/\tau_0) \to \infty$, named "run-away" [7, 8]. Here $\tau_0 = e^2/6\pi\varepsilon_0 m_0 c^3 = \mathrm{O}(10^{-24}\,\mathrm{sec})$, where $m_0$, $e$ and $\tau$ denote the rest mass, the charge and the proper time of an electron. In my previous research, I succeeded to perform the stabilization of this run-away in the QED vacuum fluctuation [8-10]. The last form of my equation was

$$\frac{dw^\mu}{d\tau} = -\frac{e}{m_0(1-\eta f_0)}[\mathfrak{F}^{\mu\nu} + \eta g_0(*\mathfrak{F})^{\mu\nu}]w_\nu \,. \tag{1}$$

The vector space $\mathbb{V}_M^4$ denotes the set of vectors in Minkowski spacetime $(\mathbb{A}^4, g)$ [I]. Defining $*\mathbb{V}_M^4$ as the dual space of $\mathbb{V}_M^4$, the Lorentz metric $g \in *\mathbb{V}_M^4 \otimes *\mathbb{V}_M^4$ has the signature of $(+,-,-,-)$, for $\forall a, b \in \mathbb{V}_M^4$, $g_{\mu\nu}a^\mu b^\nu = a^\nu b_\nu = a^0 b^0 - a^1 b^1 - a^2 b^2 - a^3 b^3 \in \mathbb{R}$. $w$ is the 4-velocity defined by $w = dx/d\tau = \gamma(c, \mathbf{v}) \in \mathbb{V}_M^4$. The field $\mathfrak{F} = F_{ex} + F_{LAD} \in \mathbb{V}_M^4 \otimes \mathbb{V}_M^4$, $F_{ex} \in \mathbb{V}_M^4 \otimes \mathbb{V}_M^4$ is an arbitrary external field, in our case generated by lasers. The field acting on an electron $F_{LAD} \in \mathbb{V}_M^4 \otimes \mathbb{V}_M^4$ is the radiation LAD field,

$$F_{LAD}^{\mu\nu}\Big|_{x=x(\tau)} = -\frac{m_0 \tau_0}{ec^2}\left(\frac{d^2 w^\mu}{d\tau^2} w^\nu - w^\mu \frac{d^2 w^\nu}{d\tau^2}\right). \tag{2}$$

Since $\eta = O(\hbar^3)$, the limit $\hbar \to 0$ in Eq.(1) derives from the equation of motion $m_0 dw^\mu/d\tau = -e\mathfrak{F}^{\mu\nu}w_\nu$, the so-called LAD equation. $f_0$ and $g_0$ are Lorentz invariant functions depending on the model of QED vacuum. In the case of the Heisenberg-Euler vacuum [11,12], $f_0 = \langle \mathfrak{F} | \mathfrak{F} \rangle = \mathfrak{F}_{\mu\nu}\mathfrak{F}^{\mu\nu}$ [II] and $g_0 = 7/4 \times \langle \mathfrak{F} | *\mathfrak{F} \rangle = 7/4 \times \mathfrak{F}_{\mu\nu}(*\mathfrak{F})^{\mu\nu}$ [8-10]. These works suggest (i) the QED vacuum fluctuation stabilizes the LAD field and (ii) it behaves well since Eq.(1) agreed with one of the major references proposed by Landau and Lifshitz [13].

On the other hand, it is considered that the dynamics of an electron should be corrected in the high-intense fields produced by 10PW lasers, by QED-based synchrotron radiation. In this physics regime, it is often discussed in terms of the parameter $\chi \in \mathbb{R}$ representing the field strength [14].

$$\chi = \frac{3}{2}\frac{\lambdabar_C}{m_0 c^2}\sqrt{-g_{\mu\nu}f_{ex}^{\ \mu}f_{ex}^{\ \nu}}, \tag{3}$$

where the Compton length $\lambdabar_C = \hbar/m_0 c$ and $r_\chi = r/(1-\chi r)$. When one considers this in the rest frame, $\chi = 3/2 \times |\mathbf{E}_{ex}|_{rest}/E_{schwinger}$. Here, $E_{Schwinger} = m_0^2 c^3/e\hbar$ is the critical field strength of light, namely the Schwinger limit. Therefore $\chi$ represents the external field strength or the intensity by using the ratio with this limit. By using QED-based synchrotron radiation with this $\chi$ dependence, I. Sokolov, et al. [14] proposed the following radiation reaction model:

$$\frac{dp^\mu}{d\tau} = -eF_{ex}^{\mu\nu}\frac{dx_\nu}{d\tau} + \frac{\tau_0 q(\chi)}{m_0^2 c^2}g_{\alpha\beta}f_{ex}^{\ \alpha}f_{ex}^{\ \beta}p^\mu \tag{4}$$

$$\frac{dx^\mu}{d\tau} = \frac{1}{m_0}p^\mu + \tau_0 q(\chi)\frac{f_{ex}^\mu}{m_0} \tag{5}$$

We will reference this as the QED-Sokolov equation/model since the function $q(\chi)$ depends on the QED cross-section of synchrotron radiation:

---

[I] $\mathbb{A}^4$ is the 4-dimensional affine space. The linear subspace of $\mathbb{A}^4$ should be $\mathbb{V}_M^4$.
[II] For $\forall A, B \in \mathbb{V}_M^4 \otimes \mathbb{V}_M^4$, $\langle A | B \rangle \equiv A_{\mu\nu}B^{\mu\nu}$.

$$q(\chi) = \frac{9\sqrt{3}}{8\pi} \int_0^{\chi^{-1}} dr\, r \left[ \int_{r_\chi}^\infty dr'\, K_{5/3}(r') + \chi^2 r r_\chi K_{2/3}(r_\chi) \right] \quad \text{III}. \tag{6}$$

The set of Eqs.(4-6) incorporates the modification of the QED radiation spectrum into the model [14]. In the low-intensity field regime, $\chi \ll 1$, then $q(\chi) \approx 1$. This limit converges to the result of the Landau-Lifshitz equation [13]. On the other hand in the case of $\chi \sim 1$, which means $10^{22}$W/cm$^2$-class laser and a GeV electron, $q(\chi) \sim 0.2$. So, the function $q(\chi)$ modifies radiation from the classical to quantum high-field dynamics. However, the QED-Sokolov equation violates the Lorentz invariant, $(dx^\mu/d\tau)(dx_\mu/d\tau) = c^2$ which should be satisfied under classical dynamics; we should recover this requirement when we consider the classical-relativistic equation of motion.

It is natural to consider that the difference of the radiation field between classical dynamics and QED is the alteration of the source (current) term in Maxwell's equation. When we consider QED effects for radiation reaction in the framework of classical dynamics, we need to insert the modulation of the charge-current density for describing the QED-based radiation field. In this paper, I discuss the general method of adapting the models from the modified radiation field $F_{\text{Mod-LAD}}$ in high-intensity external fields (such as laser) to the field propagation in QED vacuum with the new degrees of freedom as the extension from Ref.[9] and Ref.[10]. By the combination of these, we can find the anisotropy of the coupling factor $\mathcal{K} \in \mathbb{V}_M^4 \otimes \mathbb{V}_M^4 \otimes \mathbb{V}_M^4 \otimes \mathbb{V}_M^4$ between an electron and fields, which is a unique dynamics behavior predicted by this new model.

$$\left.\begin{array}{l} 1)\ F_{\text{LAD}} \mapsto F_{\text{Mod-LAD}}\ (\text{Sect.2.1}) \\ 2)\ \text{QED vacuum (Sect.2.2)} \end{array}\right\} \xRightarrow[(\text{Sect.3.3}\,/\,\text{Appendices 1})]{3)\ \Xi = q(\chi)} \quad \frac{dw^\mu}{d\tau} = -\frac{e}{m_0} \mathcal{K}^{\mu\nu}{}_{\alpha\beta}(F_{\text{ex}}{}^{\alpha\beta} + F_{\text{Mod-LAD}}{}^{\alpha\beta}) w_\nu$$

For the demonstration of this scheme, I will introduce the new functions $\Xi$ and $\Theta$ for the modification of the LAD field in the high-intensity external field at first. By using them, I will derive the modified-LAD field $F_{\text{Mod-LAD}}$ corresponding to the QED-Sokolov equation (Sect.2.1) and correct the field $F_{\text{ex}} + F_{\text{Mod-LAD}}$ by QED vacuum fluctuation (Sect.2.2). To simplify, I perform it by the field propagation in Heisenberg-Euler vacuum in Ch.3. We reach the conclusion that the new equation agrees well with the QED-Sokolov equation with the relation $(dx^\mu/d\tau)(dx_\mu/d\tau) = c^2$ and the anisotropic coupling between an electron and fields.

## 2. Modification by High-intensity field

In this chapter, I discuss the general method of how to treat the field $\widetilde{\mathfrak{F}} \in \mathbb{V}_M^4 \otimes \mathbb{V}_M^4$ acting on an electron; $\widetilde{\mathfrak{F}}^{\mu\nu} = \mathcal{K}^{\mu\nu}{}_{\alpha\beta}(F_{\text{ex}}{}^{\alpha\beta} + F_{\text{Mod-LAD}}{}^{\alpha\beta})$. By using this field, we can obtain the equation of motion of an electron.

### 2.1. Introduction of running charge and mass

In ultrahigh-intensity fields, the coupling (charge) of an electron to fields may be modified due to the alteration of the current from classical dynamics to QED. This formulation has been discussed by I. Sokolov,

---

III For $dW/dt|_{\text{classic}} = -\tau_0/m_0 \times g_{\alpha\beta} f_{\text{ex}}{}^\alpha f_{\text{ex}}{}^\beta$ as classical radiation energy loss (the Lamor's formula), the QED corrected formula becomes $dW/dt|_{\text{High Field}} = q(\chi) \times dW/dt|_{\text{classic}}$ [14].

*et al.* which was introduced above as Eqs.(3-6) [14]. Finally, they formulated the following interesting relation,

$$\tau_0|_{\text{High Field}} = q(\chi) \times \tau_0 = q(\chi) \times \frac{e^2}{6\pi\varepsilon_0 m_0 c^3} \quad . \tag{7}$$

where $\tau_0$ is the constant in Eq.(2), $\tau_0 = e^2/6\pi\varepsilon_0 m_0 c^3 = O(10^{-24}\text{ sec})$ and $c\tau_0$ describes the order of the classical electron radius. Equation (7) suggests the coupling $e^2/m_0$ should be replaced by $q(\chi) \times e^2/m_0$. It seems that Eq.(7) means the replacements of the charge $e \mapsto e' = e \times \sqrt{q(\chi)}$ and the LAD field $F_{\text{LAD}} \mapsto F'_{\text{LAD}} = \sqrt{q(\chi)} \times F_{\text{LAD}}$ since $F_{\text{LAD}} = O(e)$, but it is not correct. If we accept this replacement, $dw^\mu/d\tau = -e'/m_0 \times (F_{\text{ex}}^{\mu\nu} + F'_{\text{LAD}}^{\mu\nu})w_\nu = -e/m_0 \times [\sqrt{q(\chi)} \times F_{\text{ex}}^{\mu\nu} + q(\chi) \times F_{\text{LAD}}^{\mu\nu}]w_\nu$ and the term of the external force $-[e \times \sqrt{q(\chi)}]/m_0 \times F_{\text{ex}}^{\mu\nu} w_\nu$ appears in the equation of motion. However this term should be just $-e/m_0 \times F_{\text{ex}}^{\mu\nu} w_\nu$ for describing the incoming background field $F_{\text{ex}}$ [see in the QED-Sokolov equation (4)]. In the case of doing the replacements $e \mapsto e' = e \times q(\chi)$, $F_{\text{LAD}} \mapsto F'_{\text{LAD}} = q(\chi) \times F_{\text{LAD}}$ and $m_0 \mapsto m_0' = m_0 \times q(\chi)$, it follows that $dw^\mu/d\tau = -e/m_0 \times [F_{\text{ex}}^{\mu\nu} + q(\chi) \times F_{\text{LAD}}^{\mu\nu}]w_\nu$, which is very similar to the form of the QED-Sokolov model in Eqs.(4-5). Therefore, it requires us to put the running charge and mass for the realization of QED-based synchrotron radiation like QED-Sokolov model.

Following the above idea, I pass to a more general discussion. The requirement for the modification of radiation is that the charge and mass of an electron should be also running. We introduce the new non-zero functions $\Xi$, $\Theta \in C^\infty(\mathbb{R})$ satisfying $q(\chi) = \Xi^2/\Theta$. Then, we can find the replacements of $e \mapsto e_{\text{High Field}} = e \times \Xi$ and $m_0 \mapsto m_{\text{High Field}} = m_0 \times \Theta$ with Eq.(7), $\tau_0 = e^2/6\pi\varepsilon_0 m_0 c^3 \mapsto \tau_0|_{\text{High Field}} = e_{\text{High Field}}^2/6\pi\varepsilon_0 m_{\text{High Field}} c^3$. The two functions $\Xi$ and $\Theta$ should be the Lorentz invariants.

From here, we try to derive again the equation of radiation reaction with the running charge $e_{\text{High Field}}$ and the running mass $m_{\text{High Field}}$ under high-intensity fields and also demonstrate the relation $\Xi = \Theta = q(\chi)$ as the plausible candidate. For the realization of QED-based synchrotron radiation, we borrow the result from QED, Eq.(6). At first we consider the modification of the LAD field for adopting the QED synchrotron radiation. The equation of an electron's motion and the Maxwell equation with $e_{\text{High Field}}$ and $m_{\text{High Field}}$ become:

$$m_{\text{High Field}}(\tau)\frac{dw^\mu}{d\tau} = -e_{\text{High Field}}(\tau)\mathfrak{F}_{\text{hom}}^{\mu\nu} w_\nu \tag{8}$$

$$\partial_\mu F^{\mu\nu} = -c\mu_0 \int_{-\infty}^{\infty} d\tau' e_{\text{High Field}}(\tau') w^\nu(\tau') \delta^4(x' - x(\tau')) \tag{9}$$

Here, $\mathfrak{F}_{\text{hom}} \in \mathbb{V}_M^4 \otimes \mathbb{V}_M^4$ is the homogenous solution of Eq.(9). The solutions of Eq.(9) are the retarded and the advance field [7, 15] [IV].

---

[IV] The derivation of this field is based on Ref.[15]. By using the Green function $G_{\text{ret,adv}}(x, x')$, the solution of the Maxwell's equation (9) is, $A_{\text{ret,adv}}^\nu(x) = -ec\mu_0 \int_{-\infty}^{\infty} d\tau' \, \Xi(\tau') w^\nu(\tau') G_{\text{ret,adv}}(x, x(\tau'))$. The field Eqs.(10-11) is derived from the relation $F_{\text{ret,adv}}^{\mu\nu}(x) = -ec\mu_0 \int_{-\infty}^{\infty} d\tau' \, \Xi(\tau')[w^\nu(\tau')\partial^\mu - w^\mu(\tau')\partial^\nu] G_{\text{ret,adv}}(x, x(\tau'))$ at the point, $x = x(\tau)$.

$$F_{\text{ret}}{}^{\mu\nu}\Big|_{x=x(\tau)} = \frac{3}{4}\frac{m_0\tau_0}{ec^2}\left[\frac{d(\Xi w^\mu)}{d\tau}w^\nu - w^\mu\frac{d(\Xi w^\nu)}{d\tau}\right]\int_{-\infty}^{\infty}d\delta\tau\,\frac{\delta(\delta\tau)}{|\delta\tau|}$$

$$-\frac{m_0\tau_0}{ec^2}\left[\frac{d^2(\Xi w^\mu)}{d\tau^2}w^\nu - \frac{d^2(\Xi w^\nu)}{d\tau^2}w^\mu\right] \quad (10)$$

$$F_{\text{adv}}{}^{\mu\nu}\Big|_{x=x(\tau)} = \frac{3}{4}\frac{m_0\tau_0}{ec^2}\left[\frac{d(\Xi w^\mu)}{d\tau}w^\nu - w^\mu\frac{d(\Xi w^\nu)}{d\tau}\right]\int_{-\infty}^{\infty}d\delta\tau\,\frac{\delta(\delta\tau)}{|\delta\tau|}$$

$$+\frac{m_0\tau_0}{ec^2}\left[\frac{d^2(\Xi w^\mu)}{d\tau^2}w^\nu - \frac{d^2(\Xi w^\nu)}{d\tau^2}w^\mu\right] \quad (11)$$

Following Dirac's ideas "the radiated field $= (F_{\text{ret}}(x) - F_{\text{adv}}(x))/2$", we can obtain the modified LAD field,

$$F_{\text{Mod-LAD}}{}^{\mu\nu}\Big|_{x=x(\tau)} = -\frac{m_0\tau_0}{ec^2}\left[\frac{d^2(\Xi w^\mu)}{d\tau^2}w^\nu - \frac{d^2(\Xi w^\nu)}{d\tau^2}w^\mu\right]$$

$$= \Xi \times F_{\text{LAD}}{}^{\mu\nu}\Big|_{x=x(\tau)} - \frac{2m_0\tau_0}{ec^2}\frac{d\Xi}{d\tau}\left(\frac{dw^\mu}{d\tau}w^\nu - \frac{dw^\nu}{d\tau}w^\mu\right). \quad (12)$$

We can find that this field avoids the singularity of $\int_{-\infty}^{\infty}d\delta\tau\,\delta(\delta\tau)/|\delta\tau|$. When the factor $\Xi \to 1$, the field $F_{\text{Mod-LAD}} \to F_{\text{LAD}}$ smoothly. Defining the homogenous field $\mathfrak{F}_{\text{hom}} = F_{\text{ex}} + F_{\text{Mod-LAD}} \in \mathbb{V}_M^4 \otimes \mathbb{V}_M^4$, the equation of an electron's motion Eq.(8) becomes as follows:

$$m_0\frac{dw^\mu}{d\tau} = -e\frac{\Xi}{\Theta}\frac{1}{1-2\tau_0\frac{\Xi}{\Theta}\frac{d\Xi}{d\tau}}\mathfrak{F}_{(\Xi)}{}^{\mu\nu}w_\nu . \quad (13)$$

Here $\mathfrak{F}_{(\Xi)} = F_{\text{ex}} + \Xi \times F_{\text{LAD}} \in \mathbb{V}_M^4 \otimes \mathbb{V}_M^4$. Next, we proceed to the demonstration of the relation $\Xi = \Theta = q(\chi)$. At first, we assume this equation includes the terms of the QED-Sokolov equation. Assuming the variation of $\Xi$ is very slow, the orders of the magnitude of $\Xi$ and $\Theta$ are the same, then we obtain $\Xi/\Theta \times |\tau_0 d\Xi/d\tau| \ll 1$. This equation (13) cannot be solved by the same reason as run-away on the LAD equation due to the term of the second order derivative, so-called the Schott term[V] $m_0\tau_0/(1-2\Xi/\Theta \times \tau_0 d\Xi/d\tau)\times d^2w/d\tau^2 \in \mathbb{V}_M^4$. For estimating $\Xi$ and $\Theta$, we use the perturbation as the method by Landau-Lifshitz [13] with the definition $p = m_0 w \in \mathbb{V}_M^4$,

$$\frac{dp^\mu}{d\tau} = -\frac{e}{m_0}F_{\text{ex}}{}^{\mu\nu}\left(\frac{\Xi}{\Theta}p_\nu + \tau_0\frac{\Xi^2}{\Theta}g_{\nu\lambda}f_{\text{ex}}{}^\lambda\right) + \frac{\tau_0}{m_0{}^2c^2}\frac{\Xi^2}{\Theta}g_{\theta\lambda}f_{\text{ex}}{}^\theta f_{\text{ex}}{}^\lambda p^\mu$$

$$-\frac{e}{m_0}\tau_0\frac{\Xi^2}{\Theta}\frac{dF_{\text{ex}}{}^{\mu\theta}}{d\tau}p_\theta - 2\frac{e}{m_0}\tau_0\frac{\Xi^2}{\Theta^2}\frac{d\Xi}{d\tau}F_{\text{ex}}{}^{\mu\nu}p_\nu + O(\tau_0{}^2) . \quad (14)$$

Where we can find Eq.(7), $\tau_0|_{\text{High Field}} = \Xi^2/\Theta \times \tau_0 = q(\chi) \times \tau_0$ and the direct radiation term $\tau_0 q(\chi)/m_0 \times$ $g_{\alpha\beta} f_{ex}{}^\alpha f_{ex}{}^\beta p^\mu$ [VI] in Eq.(14). This direct radiation term also appears in the QED-Sokolov equation (4). For fitting the QED-Sokolov model as mentioned at the beginning of this section,

$$\Xi = \Theta \quad (15)$$

is required since $-e/m_0 \times F_{ex}{}^{\mu\nu}[\Xi/\Theta \times p_\nu + \tau_0 q(\chi) g_{\nu\lambda} f_{ex}{}^\lambda] = -e/m_0 F_{ex}{}^{\mu\nu}[p_\nu + \tau_0 q(\chi) g_{\nu\lambda} f_{ex}{}^\lambda]$ should be satisfied. The final term in the LHS in Eq.(14) vanishes since $\Xi^2/\Theta^2 \times |\tau_0 d\Xi/d\tau| = |\tau_0 dq(\chi)/d\tau| \ll 1$, the difference between Eq.(14) and the QED-Sokolov model is just $-e/m_0 \times [\tau_0 q(\chi) dF_{ex}{}^{\mu\theta}/d\tau] p_\theta$. However, we know this term (the approximation of the Schott term in the LAD equation) also vanished in many numerical tests. Therefore, the relation $\Xi = \Theta = q(\chi)$ should be satisfied for describing QED-based synchrotron radiation in the equation of motion. Inserting these relations, Eq.(13) becomes

$$m_0 \frac{dw^\mu}{d\tau} = -e \frac{1}{1 - 2\tau_0 \frac{d\Xi}{d\tau}} \mathfrak{F}_{(\Xi)}{}^{\mu\nu} w_\nu . \quad \text{[VII]} \quad (16)$$

---

[V] The Schott term in the LAD equation is $m_0\tau_0 d^2w/d\tau^2 \in \mathbb{V}_M^4$.

Equation (16) is one of the methods for radiation reaction with QED synchrotron radiation, however it suffers from the run-away problem. I also present the method of stabilizing the singularity of the field $\mathfrak{F} = F_{ex} + F_{\text{Mod-LAD}}$ before considering the equation of motion in the next section.

## 2.2. Stabilization by QED vacuum fluctuation

In Sect.2.1, I modified the LAD field by introducing the running charge $e_{\text{High Field}} = e \times \Xi$, to obtain the modified LAD field $F_{\text{Mod-LAD}} \in \mathbb{V}_M^4 \otimes \mathbb{V}_M^4$. In the following section, we consider how to stabilize the field $\mathfrak{F}_{\text{hom}} = F_{ex} + F_{\text{Mod-LAD}} \in \mathbb{V}_M^4 \otimes \mathbb{V}_M^4 \in \mathbb{V}_M^4 \otimes \mathbb{V}_M^4$ which is the homogenous solution of the source-free Maxwell's equation (9). At first, the field $F_{\text{Mod-LAD}}$ satisfies the source free Maxwell's equation $\partial_\mu F_{\text{Mod-LAD}}{}^{\mu\nu} = 0$ [VIII]. Replacing $F_{\text{LAD}}$ by $F_{\text{Mod-LAD}}$ under the method of Ref.[10], we can find the homogenous field $\mathfrak{F}_{\text{hom}} = F_{ex} + F_{\text{Mod-LAD}}$ [IX] at the observation point far from an electron. The field dresses the vacuum polarization during the field propagation, $\mathfrak{F}_{\text{hom}}$ represents the already "dressed" field [8-10]. Here we need to derive the undressed field $\mathfrak{F} \in \mathbb{V}_M^4 \otimes \mathbb{V}_M^4$ acting on an electron for substituting into Eq.(8). The general dynamics of the propagating field is described by

$$L(\langle \mathfrak{F} | \mathfrak{F} \rangle, \langle \mathfrak{F} | *\mathfrak{F} \rangle) = -\frac{1}{4\mu_0} \langle \mathfrak{F} | \mathfrak{F} \rangle + L_{\text{Quantum Vacuum}}(\langle \mathfrak{F} | \mathfrak{F} \rangle, \langle \mathfrak{F} | *\mathfrak{F} \rangle) . \quad (17)$$

Here, $L_{\text{Quantum Vacuum}}$ is an undefined Lagrangian density for the QED vacuum fluctuation. The important remark is that this Lagrangian density is applicable only to describe the field propagation in the spacetime

---

[VI] The direct radiation term in the LAD equation is $m_0 \tau_0 g_{\alpha\beta}(dw^\alpha/d\tau)(dw^\beta/d\tau) w \in \mathbb{V}_M^4$.

[VII] Eq.(16) derive $dp^\mu/d\tau = -e/m_0 \times F_{ex}{}^{\mu\nu}[p_\nu + \tau_0 q(\chi) g_{\nu\lambda} f_{ex}{}^\lambda] + \tau_0 q(\chi)/m_0^2 c^2 \times g_{\theta\lambda} f_{ex}{}^\theta f_{ex}{}^\lambda p^\mu$
$- e/m_0 \times \tau_0 q(\chi) dF_{ex}{}^{\mu\theta}/d\tau \, p_\theta - 2e/m_0 \times \tau_0 \, dq(\chi)/d\tau \, F_{ex}{}^{\mu\nu} p_\nu + O(\tau_0^2)$, the quasi-QED-Sokolov equation.

[VIII] Denoting $\partial_\mu F_{\text{ret,adv}}{}^{\mu\nu} = \mu_0[-ec \int_{-\infty}^{\infty} d\tau' \Xi(\tau') w^\nu(\tau') \delta^4(x' - x(\tau'))]$ and $F_{\text{Mod-LAD}}{}^{\mu\nu} = (F_{\text{ret}}{}^{\mu\nu} - F_{\text{adv}}{}^{\mu\nu})/2$.

[IX] When $\partial_\mu F_{ex}{}^{\mu\nu} = 0$, the followings are satisfied: $\langle F_{ex} | F_{ex} \rangle = 0$ and $\langle F_{ex} | *F_{ex} \rangle = 0$.

6
Keita Seto, "Radiation Reaction in High-Intense Fields",
Final-Submissionn to PTEP: August 21, 2015.

without any field sources. By solving this, we can obtain the following Maxwell's equation:

$$\partial_\mu \left[ \mathfrak{F}^{\mu\nu} + \frac{1}{c\varepsilon_0} M^{\mu\nu} \right] = 0 \qquad (18)$$

Equation (18) it the Maxwell's equation for the source-free field, $\mathfrak{F} + M/c\varepsilon_0$. $M \in \mathbb{V}_M^4 \otimes \mathbb{V}_M^4$ being the polarization of vacuum [9,10],

$$\frac{1}{c\varepsilon_0} M^{\mu\nu} = -\eta f \times \mathfrak{F}^{\mu\nu} - \eta g \times {}^*\mathfrak{F}^{\mu\nu} \qquad (19)$$

$$\eta f\left(\langle \mathfrak{F} | \mathfrak{F} \rangle, \langle \mathfrak{F} | {}^*\mathfrak{F} \rangle\right) = 4\mu_0 \frac{\partial L_{\text{Quantum Vacuum}}}{\partial \langle \mathfrak{F} | \mathfrak{F} \rangle} \qquad (20)$$

$$\eta g\left(\langle \mathfrak{F} | \mathfrak{F} \rangle, \langle \mathfrak{F} | {}^*\mathfrak{F} \rangle\right) = 4\mu_0 \frac{\partial L_{\text{Quantum Vacuum}}}{\partial \langle \mathfrak{F} | \mathfrak{F} \rangle} \qquad (21)$$

Here, $\eta = 4\alpha^2 \hbar^3 \varepsilon_0 / 45 m_0^4 c^3$. $\mathfrak{F} + M/c\varepsilon_0$ refers to the dressed field set of $(\mathbf{D}, \mathbf{H})$. In addition, the following Maxwell's equation is also held: $\partial_\mu \mathfrak{F}_{\text{hom}}^{\mu\nu} = 0$. Thus, the solution of Eq.(18), $\mathfrak{F} + M/c\varepsilon_0$ connects to $(\mathbf{D}, \mathbf{H}) = \mathfrak{F}_{\text{hom}} = F_{\text{ex}} + F_{\text{Mod-LAD}}$ with the continuity and smoothness with $C^\infty$ at all points in the Minkowski spacetime,

$$\boxed{\mathfrak{F}^{\mu\nu} - \eta f \times \mathfrak{F}^{\mu\nu} - \eta g \times {}^*\mathfrak{F}^{\mu\nu} = \mathfrak{F}_{\text{hom}}^{\mu\nu}} \ . \qquad (22)$$

Via the algebraic treatments, we can solve Eq.(22) for $\mathfrak{F}$,

$$\mathfrak{F}^{\mu\nu} = \frac{(1-\eta f)\mathfrak{F}_{\text{hom}}^{\mu\nu} + \eta g({}^*\mathfrak{F}_{\text{hom}})^{\mu\nu}}{(1-\eta f)^2 + (\eta g)^2} \qquad (23)$$

(see Ref.[10]). We propose the following equation of motion for a radiating electron in the high-intensity fields coupling with the Maxwell's equation (18).

$$m_{\text{High Field}}(\tau) \frac{dw^\mu}{d\tau} = -e_{\text{High Field}}(\tau) \mathfrak{F}^{\mu\nu} w_\nu \ . \qquad (24)$$

Rewriting Eq.(24) with the relation $\Xi = \Theta$ and following Sect. 2-1, we can get the form,

$$\boxed{\frac{dw^\mu}{d\tau} = -\frac{e}{m_0} \frac{[(1-\eta f_0)\mathfrak{F}_{\text{hom}}^{\mu\nu} + \eta g_0({}^*\mathfrak{F}_{\text{hom}})^{\mu\nu}]}{(1-\eta f_0)^2 + (\eta g_0)^2} w_\nu} \qquad (25)$$

(see Appendix 1). Here, $f_0 = f(\langle \mathfrak{F}_{\text{hom}} | \mathfrak{F}_{\text{hom}} \rangle, \langle \mathfrak{F}_{\text{hom}} | {}^*\mathfrak{F}_{\text{hom}} \rangle)$ and $g_0 = g(\langle \mathfrak{F}_{\text{hom}} | \mathfrak{F}_{\text{hom}} \rangle, \langle \mathfrak{F}_{\text{hom}} | {}^*\mathfrak{F}_{\text{hom}} \rangle)$. Introducing the new tensor

$$\mathcal{K}^{\mu\nu\alpha\beta} = \frac{(1-\eta f_0) \times g^{\mu\alpha} g^{\nu\beta} + \eta g_0 \times \frac{1}{2!} \varepsilon^{\mu\nu\alpha\beta}}{(1-\eta f_0)^2 + (\eta g_0)^2} \ , \qquad (26)$$

the equation of an electron's motion becomes briefly,

$$\boxed{\frac{dw^\mu}{d\tau} = -\frac{e}{m_0}\mathfrak{K}^{\mu\nu}{}_{\alpha\beta}\mathfrak{F}_{\text{hom}}{}^{\alpha\beta}w_\nu} \quad . \tag{27}$$

For the mimic of the Sokolov's model, the function $\Xi$ should have the dependence $\Xi \to 1$ in the low-intense limit converging to Eq.(1).

## 3. High-intensity field correction under the first order Heisenberg-Euler vacuum

In this chapter, we consider Eq.(25) or Eq.(27) with the Heisenberg-Euler model for the QED vacuum fluctuation. After the derivation of the equation of an electron's motion, we demonstrate the stability of this equation and perform the numerical calculations of it. We choose the relation $\Theta = \Xi = q(\chi)$ in this chapter, however we describe it by using $\Xi$ for the extension in Sect.3.1-3.2.

### 3.1 Equation of motion

The familiar model of QED vacuum was represented by Heisenberg and Euler [11, 12]:

$$L_{\text{Quantum Vacuum}} = L_{\substack{\text{the lowest order of}\\ \text{Heisenberg-Euler}}} = \frac{\alpha^2 \hbar^3 \varepsilon_0^2}{360 m_0^4 c}\left[4\langle\mathfrak{F}|\mathfrak{F}\rangle^2 + 7\langle\mathfrak{F}|*\mathfrak{F}\rangle^2\right] \tag{28}$$

The Heisenberg-Euler Lagrangian basically presents the dynamics only for the constant field. If more general Lagrangian for any fields exists, that generalized Lagrangian includes the component of Eq.(28) since the constant field should be one of the behaviors of the general Lagrangian. In this section, I assume we can apply Eq.(28) for the field propagation like in Ref.[10]. In this case, the functions $f_0$ and $g_0$ are

$$f_0 = \langle\mathfrak{F}_{\text{hom}}|\mathfrak{F}_{\text{hom}}\rangle = \langle F_{\text{Mod-LAD}}|F_{\text{Mod-LAD}}\rangle + 2\times\langle F_{\text{Mod-LAD}}|F_{\text{ex}}\rangle \tag{29}$$

$$g_0 = \frac{7}{4}\langle\mathfrak{F}_{\text{hom}}|*\mathfrak{F}_{\text{hom}}\rangle = \frac{7}{2}\langle F_{\text{Mod-LAD}}|*F_{\text{ex}}\rangle \tag{30}$$

and we transform Eq.(25) like:

$$\frac{dw^\mu}{d\tau} = -\frac{e}{m_0}\frac{(1-\eta\langle\mathfrak{F}_{\text{hom}}|\mathfrak{F}_{\text{hom}}\rangle)\mathfrak{F}_{\text{hom}}{}^{\mu\nu} + \frac{7}{4}\eta\langle\mathfrak{F}_{\text{hom}}|*\mathfrak{F}_{\text{hom}}\rangle(*\mathfrak{F}_{\text{hom}})^{\mu\nu}}{(1-\eta\langle\mathfrak{F}_{\text{hom}}|\mathfrak{F}_{\text{hom}}\rangle)^2 + \left(\frac{7}{4}\eta\langle\mathfrak{F}_{\text{hom}}|*\mathfrak{F}_{\text{hom}}\rangle\right)^2}w_\nu \quad . \tag{31}$$

We can find the singularity when $\eta g_0 = 0$ and $1-\eta f_0 = 1-\eta\langle\mathfrak{F}_{\text{hom}}|\mathfrak{F}_{\text{hom}}\rangle = 0$ in Eq.(31). From the condition of the low-intense limit, $1-\eta f_0 > 0$ must be required for avoiding the singular point:

$$1-\eta f_0 = \frac{2\eta}{e^2 c^2} \times [m_0 \tau_0 (\Xi \ddot{\mathbf{v}} + 2\dot{\Xi}\dot{\mathbf{v}}) - e\mathbf{E}_{ex}]^2 \Big|_{rest} + 1 - \frac{2\eta \mathbf{E}_{ex}^2}{c^2}\Big|_{rest} > 1 - \frac{2\eta \mathbf{E}_{ex}^2}{c^2}\Big|_{rest} \overset{\text{physical requirements}}{>} 0 \quad \text{X} \quad (32)$$

Where I employed the relation $\langle F_{\text{Mod-LAD}} | F_{\text{Mod-LAD}} \rangle = -2(m_0\tau_0/ec)^2 \times (\Xi\ddot{\mathbf{v}} + 2\dot{\Xi}\dot{\mathbf{v}})^2 |_{rest} \leq 0$ and $\langle F_{\text{Mod-LAD}} | F_{ex} \rangle = 2m_0\tau_0/ec^2 \times (\Xi\ddot{\mathbf{v}} + 2\dot{\Xi}\dot{\mathbf{v}})\cdot \mathbf{E}_{ex}|_{rest}$. $2\eta \mathbf{E}_{ex}^2/c^2|_{rest} = (5.2 \times 10^{-5}) \times (\mathbf{E}_{ex}/E_{\text{Schwinger}})^2|_{rest}$ with the definition of the Schwinger limit field $E_{\text{Schwinger}} = m_0^2 c^3/e\hbar$. Normally, the relation $|\mathbf{E}_{ex}| << E_{\text{Schwinger}}$ is satisfied. Considering the extreme condition like $|\mathbf{E}_{ex}| = O(E_{\text{Schwinger}})$, the coefficient of $(\mathbf{E}_{ex}|_{rest}/E_{\text{Schwinger}})^2$ is smaller than unity, $1-\eta f_0 > 0$ should be held, it is the requirement for avoiding the instability and for taking into consideration the high-intense fields.

### 3.2 Run-away avoidance

In the original model of radiation reaction, the LAD equation has an instability named the "run-away" solution diverging exponentially even in the absence of an external field. This mathematical problem is also called "self-acceleration". The new equation must be required to hold the stability and we need to understand for how large dynamical range we can apply it. We assume the condition of Eq.(32) in following analysis. For instance, I rewrite the equation of an electron's motion Eq.(25/31) like

$$m_0 \frac{dw^\mu}{d\tau} = \frac{(1-\eta f_0) \times [f_{ex}^{\ \mu} - eF_{\text{Mod-LAD}}^{\ \mu\nu} w_\nu] + \eta g_0 (*f_{ex})^\mu}{(1-\eta f_0)^2 + (\eta g_0)^2}, \quad (33)$$

with the definition of the forces $f_{ex}^{\ \mu} = -eF_{ex}^{\ \mu\nu}w_\nu$ and $(*f_{ex})^\mu = -e(*F_{ex})^{\mu\nu}w_\nu$. Here, we follow the two-stage analysis used in Ref.[10]. At first, we check the finiteness of the radiation energy due to the possibility that run-away comes from infinite radiated energy. In the second stage, we proceed to the asymptotic analysis proposed by F. Röhrlich for investigation after releasing from the external field [16].

In the first stage, we make the modified-Larmor's formula $dW/d\tau = -m_0\tau_0 \Xi g_{\alpha\beta}(dw^\alpha/d\tau)(dw^\beta/d\tau)$ by the replacement $\tau_0 \mapsto \Xi \times \tau_0$ [XI] for the estimation of radiation power from Eq.(33):

$$m_0\tau_0 \Xi g_{\mu\nu} \frac{dw^\mu}{d\tau}\frac{dw^\nu}{d\tau} = \frac{\tau_0 \Xi}{m_0} \frac{\frac{e^2c^2}{2\eta}(1-\eta f_0)^2 \eta f_0 + \frac{2e^2c^2}{7\eta}(1-\eta f_0)(\eta g_0)^2}{[(1-\eta f_0)^2 + (\eta g_0)^2]^2}$$

$$+ \frac{\tau_0 \Xi}{m_0} \frac{g_{\mu\nu}[(1-\eta f_0)f_{ex}^{\ \mu} + \eta g_0 * f_{ex}^{\ \mu}][(1-\eta f_0)f_{ex}^{\ \nu} + \eta g_0 * f_{ex}^{\ \nu}]}{[(1-\eta f_0)^2 + (\eta g_0)^2]^2} \quad (34)$$

Considering invariances in the rest frame, $f_0 = -2(m_0\tau_0/ec)^2 \times [m_0\tau_0(\Xi\ddot{\mathbf{v}} + 2\dot{\Xi}\dot{\mathbf{v}}) - e\mathbf{E}_{ex}]^2|_{rest} + 2\eta \mathbf{E}_{ex}^2/c^2 = O(\Xi^2 \ddot{\mathbf{v}}_{rest}^2)$ and $g_0 = 7m_0\tau_0/ec \times (\Xi\ddot{\mathbf{v}} + 2\dot{\Xi}\dot{\mathbf{v}})\cdot\mathbf{B}_{ex}|_{rest} = O(\Xi\ddot{\mathbf{v}}_{rest})$. When the dynamics becomes run-away by infinite energy emission by light, $|\Xi\ddot{\mathbf{v}}_{rest}| \to \infty$. In the run-away case, $O(|g_0|) < O(|f_0|)$ is obviously satisfied. We use this relation under the condition of Eq.(32), Eq.(34) (the detail is in Appendix

---
[X] The subscript of "rest" means the values in the rest frame.

[XI] Of course, the well-known Larmor's formula is $dW/d\tau = -m_0\tau_0 g_{\alpha\beta}(dw^\alpha/d\tau)(dw^\beta/d\tau)$.

3):

$$\left| m_0 \tau_0 \Xi g_{\mu\nu} \frac{dw^\mu}{d\tau} \frac{dw^\nu}{d\tau} \right|$$

$$\overset{\text{run-away}}{<} \frac{\tau_0 \Xi}{m_0} \frac{e^2 c^2}{2\eta} \frac{|\eta f_0|}{(1-\eta f_0)^2} + \frac{\tau_0 \Xi}{m_0} \frac{2e^2 c^2}{7\eta} \frac{1}{1-\eta f_0} \frac{(\eta f_0)^2}{(1-\eta f_0)^2}$$

$$+ \frac{\tau_0 \Xi}{m_0} \frac{|g_{\mu\nu} f_{\text{ex}}{}^\mu f_{\text{ex}}{}^\nu|}{(1-\eta f_0)^2} + \frac{2\tau_0 \Xi}{m_0} \frac{|\eta f_0|}{1-\eta f_0} \times \frac{|g_{\mu\nu} f_{\text{ex}}{}^\mu (*f_{\text{ex}})^\nu|}{(1-\eta f_0)^2}$$

$$+ \frac{\tau_0 \Xi}{m_0} \frac{|\eta f_0|}{(1-\eta f_0)^2} \frac{|g_{\mu\nu} f_{\text{ex}}{}^\mu (*f_{\text{ex}})^\nu|}{1-\eta f_0} \tag{35}$$

The functions $1/|1-x|$, $1/|1-x|^2$, $|x|/|1-x|^2$ and $|x|^2/|1-x|^2$ are finite in the domain $x \in (-\infty, 1)$, When we choose $x = \eta f_0 \leq 2\eta \mathbf{E}_{\text{ex}}{}^2 |_{\text{rest}}/c^2 < O(10^{-5})$ below the Schwinger limit, $x$ is included in this domain. In these conditions,

$$\frac{dW}{dt} = -m_0 \tau_0 \Xi g_{\mu\nu} \frac{dw^\mu}{d\tau} \frac{dw^\nu}{d\tau} \overset{\text{run-away}}{<} \infty. \tag{36}$$

As such, the possibility for the run-away due to the infinite energy emission of light was avoided in Eq.(36).

Next, we proceed to the asymptotic analysis. For this analysis, we need to take the pre-acceleration form. The form of Eq.(33) is as follows:

$$m_0 \frac{dw^\mu}{d\tau}(\tau) = e^{\int_{\tau_a}^\tau d\tau'' \frac{1}{\Xi \tau_0}} \int_\tau^\infty \frac{d\tau'}{\tau_0} \left[ \frac{(1-\eta f_0) f_{\text{ex}}{}^\mu + \eta g_0 (*f_{\text{ex}})^\mu}{+(1-\eta f_0) \frac{m_0 \tau_0 \Xi}{c^2} g_{\alpha\beta} \frac{dw^\alpha}{d\tau} \frac{dw^\beta}{d\tau} w^\mu}{(1-\eta f_0)\Xi} \right](\tau')$$

$$\times e^{-\int_{\tau_a}^{\tau'} d\tau'' \frac{1}{\Xi \tau_0}} \times e^{-2[\ln \Xi(\tau) - \ln \Xi(\tau')]} \times e^{-\int_{\tau'}^\tau d\tau'' \frac{\eta f_0}{\Xi \tau_0}} \times e^{\int_{\tau'}^\tau d\tau'' \frac{(\eta g_0)^2}{(1-\eta f_0)\Xi \tau_0}} \tag{37}$$

Here, we employed the parameter $\tau_a < \tau$. Now we consider the acceleration $dw/d\tau$ at the infinite future, $\tau \to \infty$. Following the Röhrlich's method, the acceleration converges to zero when the external field vanishes at $\tau \to \infty$. In this limit, the dynamics becomes the classical limit $\Xi \to 1$ due to the absence of the field ($\chi = 0$), $\lim_{\chi \to 0} q(\chi) = 1$. Therefore, we can obtain the limit of Eq.(37),

$$m_0 \frac{dw^\mu}{d\tau}(\infty) = f_{\text{ex}}{}^\mu(\infty) + \frac{\eta g_0}{1-\eta f_0} \times (*f_{\text{ex}})^\mu(\infty) + \frac{m_0 \tau_0 \Xi}{c^2} g_{\alpha\beta} \frac{dw^\alpha}{d\tau} \frac{dw^\beta}{d\tau} w^\mu(\infty)$$

$$= \frac{m_0 \tau_0 \Xi}{c^2} g_{\alpha\beta} \frac{dw^\alpha}{d\tau} \frac{dw^\beta}{d\tau} w^\mu(\infty), \tag{38}$$

by using the l'Hôpital's rule at $\tau \to \infty$ [XII]. The finiteness of Eq.(36) is important for the constant velocity, otherwise $dw/d\tau(\infty) = \infty$. After this procedure, we can follow the same way of Ref.[10]. Finally, we can get the limit of the acceleration of an electron;

$$\lim_{\tau \to 0} \frac{dw^\mu}{d\tau} = 0 \tag{39}$$

This is just the requirement for the avoidance of run-away proposed by Rörhlich [16]. My model also satisfies $dw/d\tau(\infty) = 0$ after releasing from the external field. In the above we could demonstrate that the new new Eq.(33) doesn't become run-away under the condition of Eq.(32).

### 3.3 Calculations

Finally, we present the numerical calculation results with other radiation reaction models. Employing the relation $\Xi = q(\chi)$, Eq.(33) becomes

$$m_0 \frac{dw^\mu}{d\tau} = \frac{(1-\eta f_0) \times [f_{ex}{}^\mu + q(\chi) \times f_{LAD}{}^\mu] + \eta g_0 (*f_{ex})^\mu}{(1-\eta f_0)^2 + (\eta g_0)^2 - 2\tau_0 (1-\eta f_0)\frac{dq(\chi)}{d\tau}}, \tag{40}$$

where $\mathcal{F}_{q(\chi)} = F_{ex} + q(\chi)F_{LAD}$. We performed the calculation of Eq.(40) with the following models: Seto I model which is Eq.(1) [9], the Landau-Lifshitz model [13], Classical Sokolov [17] and QED-Sokolov [14]. And we name Eqs.(25/33/40) as "Seto II". I assumed the case of the head-on collision between the laser photons and an electron as the initial configuration of the simulations (Fig.1). We used the parameters of Extreme Light Infrastructure - Nuclear Physics (ELI-NP) [5-6]. The peak intensity of the laser is $1 \times 10^{22}$ W/cm$^2$ in a Gaussian shaped plane-wave like Eq.(28,29) in Ref.[10]. The pulse width is 22fsec and the laser wavelength is $0.82\mu m$. The electric field is situated in the $y$ direction, the magnetic field is in the $z$ direction. The single electron travels in the negative $x$ direction, with the energy of 600MeV initially. The numerical calculations were carried out in the laboratory frame.

The time evolution of an electron's energy shows the typical behavior of radiation reaction, as. shown in Fig.2. The energy of an electron drops from the initial energy of 600MeV. Depending on the models, the final energies of the electron converges to two separated levels. The first group includes Seto I, Landau-Lifshitz, and Classical Sokolov models near 165MeV. The second group is QED-Sokolov and Seto II models, stating around 260MeV. The difference between these two groups depends on the function $q(\chi)$, obviously. In this laser intensity and energy of an electron, $\chi$ runs from 0 to 0.3 in this case. Figure 3 presents the graph of $q(\chi)$.

---

[XII] For any function $f$, $\lim_{\tau \to \infty} \int_\tau^\infty \frac{d\tau'}{\tau_0} f(\tau') \bigg/ e^{-\int_{\tau_a}^\tau d\tau'' \frac{1}{\Xi \tau_0}} = \lim_{\tau \to \infty} f(\tau)$.

$$\begin{Bmatrix} \mathbf{E}_{\text{laser}} \\ \mathbf{B}_{\text{laser}} \end{Bmatrix} = \begin{Bmatrix} \hat{\mathbf{y}} \\ \hat{\mathbf{z}} \end{Bmatrix} \hat{\mathbf{y}} E_0 \times \exp\left[-\frac{y^2 + z^2}{(\Delta x_{\text{laser spot}})^2}\right]$$
$$\times \exp\left[-\frac{(\omega_{\text{laser}} t - k_{\text{laser}} x)^2}{(\omega_{\text{laser}} \Delta \tau)^2}\right] \times \sin(\omega_{\text{laser}} t - k_{\text{laser}} x)$$

Fig.1  Setup of the laser - electron "head-on collision". The laser photons propagate along the $x$ axis. An electron travels in the negative $x$ direction.

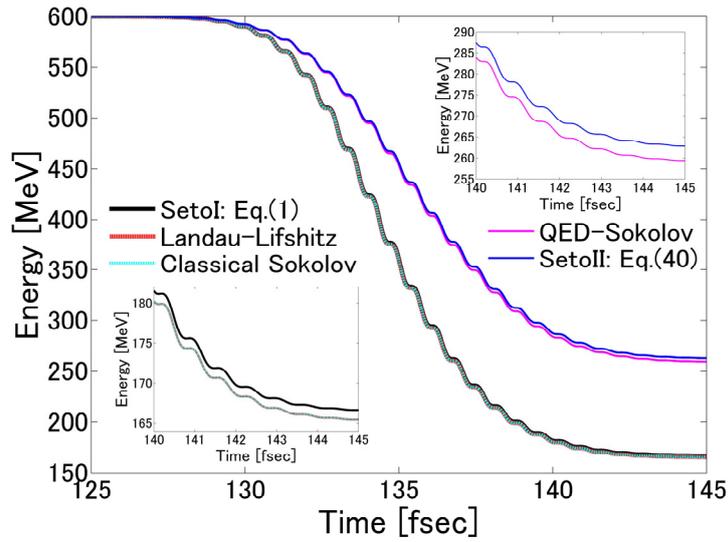

Fig.2  The time evolution of an electron's energy. The final electron's energies are, Seto I, Eq.(1): 166.5MeV, Landau-Lifshitz: 165.3MeV, Classical Sokolov: 165.3MeV, QED-Sokolov: 259.0MeV and and Seto II, Eq.(40): 262.5MeV. The insets are close-up of the figures.

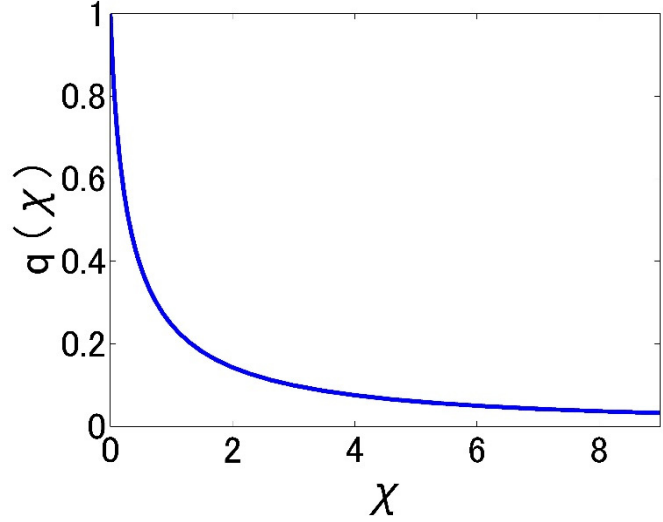

Fig.3    The function of $q(\chi)$. In this calculations, the domain is $\chi \in [0,1]$.

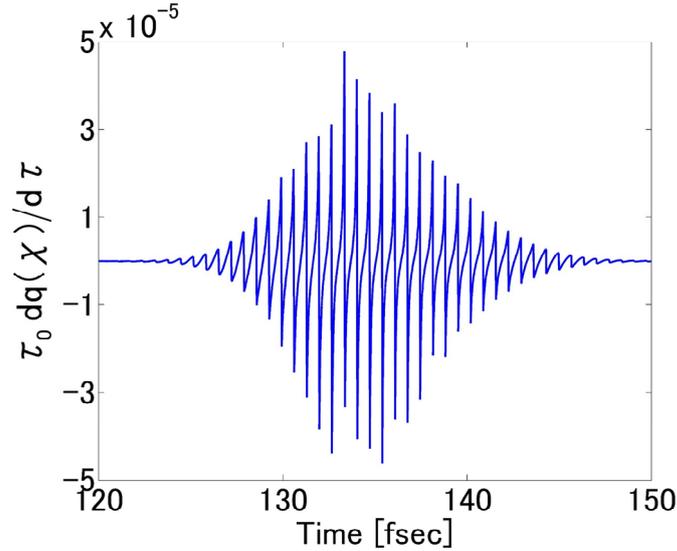

Fig.4    The function of $\tau_0 d\Xi/d\tau$.

The following is satisfied: $\tau_0 dq(\chi)/d\tau = O(10^{-5})$ (see Fig.4) and also $\eta f_0 = \eta \langle \mathfrak{F}_{hom} | \mathfrak{F}_{hom} \rangle = O(10^{-8})$. Therefore, $1 - \eta f_0 - 2\tau_0 dq(\chi)/d\tau \sim 1$ in this case. In the head-on collision between the laser photon and the electron, $\eta g_0 = 7/4 \times \eta \langle \mathfrak{F}_{hom} |{*}\mathfrak{F}_{hom} \rangle = 7 m_0 \tau_0 q(\chi)/ec \times \ddot{\mathbf{v}} \cdot \mathbf{B}_{ex}|_{rest} = 0$ since the initial condition limits the electron's motion on the $x$-$y$ plane and $\mathbf{B}_{ex}$ in the $z$ direction. Rounding the

invariance $\eta f_0$ into 1, Eq.(40) is reduced as follows:

$$m_0 \frac{dw^\mu}{d\tau} = -e \frac{1}{1 - \eta \langle \mathfrak{F}_{\text{hom}} | \mathfrak{F}_{\text{hom}} \rangle - 2\tau_0 \frac{dq(\chi)}{d\tau}} \widetilde{\mathfrak{F}}_{q(\chi)}{}^{\mu\nu} w_\nu \sim -e \frac{1}{1 - 2\tau_0 \frac{dq(\chi)}{d\tau}} \widetilde{\mathfrak{F}}_{q(\chi)}{}^{\mu\nu} w_\nu. \tag{41}$$

Since it is valid Eq.(16), we can derive the quasi-QED-Sokolov equation by using perturbation[XIII] from Eq.(40). The convergence between the model of Seto II: Eq.(40) and QED-Sokolov Eq.(4-6) appeared because of this reason. The difference of the final energies between the two groups depends on the invariant function $\Xi = q(\chi)$, $I_{\text{QED}} = q(\chi) \times I_{\text{classic}} \leq I_{\text{classic}}$ [XIV]. From the theoretical point of view, the new equation (40) can satisfy both of the relation $(dx^\mu/d\tau)(dx_\mu/d\tau) = c^2$ and $p_\mu p^\mu = m_0^2 c^2$ [XV] at any time. On the other hand, in the Classical/QED-Sokolov model, $(dx^\mu/d\tau)(dx_\mu/d\tau) \neq c^2$ and $p_\mu p^\mu = m_0^2 c^2$ [see Eq.(4)]. This is an algebraic difference between two models.

### 4. Conclusion and discussion

In summary, I updated my previous equation of a radiating electron's motion by considering the high-intensity fields and QED vacuum fluctuation. The field $\widetilde{\mathfrak{F}}$ acting on an electron was modified by the following method:

$$\left. \begin{array}{l} 1) \ F_{\text{LAD}} \mapsto F_{\text{Mod-LAD}} \ (\text{Sect.2.1}) \\ 2) \ \text{QED vacuum (Sect.2.2)} \end{array} \right\} \quad \begin{array}{c} 3) \ \Xi = q(\chi) \\ \Longrightarrow \\ (\text{Sect.3.3 / Appendices 1}) \end{array} \quad \frac{dw^\mu}{d\tau} = -\frac{e}{m_0} \mathcal{K}^\mu{}_{\alpha\beta} (F_{\text{ex}}{}^{\alpha\beta} + F_{\text{Mod-LAD}}{}^{\alpha\beta}) w_\nu$$

The external high-intensity fields modify the emitted field from an electron. The QED vacuum fluctuation stabilized the "run-away". The mathematical treatments in the derivation of the new equation was based on our previous paper [9, 10]. At first, I assumed the parameter replacements $e \mapsto e_{\text{High Field}} = e \times \Xi$ and $m_0 \mapsto m_{\text{High Field}} = m_0 \times \Theta$ in high-intensity field for taking the QED-intensity correction into the formula. In the low-intensity limit, the invariants satisfy $\Xi, \Theta \to 1$. The source term of the Maxwell's equation was deformed, depending on this replacement, the LAD field was modified from $F_{\text{LAD}}$ to $F_{\text{Mod-LAD}}$ (Sect.2.1). The field $\widetilde{\mathfrak{F}}$ which acts on an electron in the QED vacuum fluctuation should satisfy the following equation,

$$\boxed{\widetilde{\mathfrak{F}}^{\mu\nu} - \eta f \times \widetilde{\mathfrak{F}}^{\mu\nu} - \eta g \times {}^*\widetilde{\mathfrak{F}}^{\mu\nu} = F_{\text{ex}}{}^{\mu\nu} + F_{\text{Mod-LAD}}{}^{\mu\nu}}, \tag{42}$$

Then we get the following new equation of motion of an electron including radiation reaction,

---

[XIII] The "perturbation" means the replacements $dw/d\tau \mapsto f_{\text{ex}}/m_0$ in the RHS of Eq.(40) [13].

[XIV] $I_{\text{classical}} = -dW/d\tau|_{\text{classical}} = -m_0 \tau_0 g_{\alpha\beta} (dw^\alpha/d\tau)(dw^\beta/d\tau)$

[XV] $p = m_0 w = m_0 \, dx/d\tau \in \mathbb{V}_M^4$ for my new model.

$$\boxed{\frac{dw^\mu}{d\tau} = -\frac{e}{m_0} \mathcal{K}^{\mu\nu}{}_{\alpha\beta} \mathfrak{F}_{\text{hom}}{}^{\alpha\beta} w_\nu} \quad (43)$$

or the explicit form,

$$\frac{dw^\mu}{d\tau} = -\frac{e}{m_0} \frac{[(1-\eta f_0)\mathfrak{F}_{\text{hom}}{}^{\mu\nu} + \eta g_0 (*\mathfrak{F}_{\text{hom}})^{\mu\nu}]w_\nu}{(1-\eta f_0)^2 + (\eta g_0)^2} \quad (44)$$

Here, the definition of the homogeneous field $\mathfrak{F}_{\text{hom}}$ is $\mathfrak{F}_{\text{hom}} = F_{\text{ex}} + F_{\text{Mod-LAD}}$ (Sect.2.2). Here, we employed the relation $\Xi = \Theta = q(\chi)$ from QED-based synchrotron radiation. From the analysis, the following relation must be satisfied for the stability of this equation in Heisenberg-Euler vacuum [11,12]: $1 - 2\eta \mathbf{E}_{\text{ex}}^2|_{\text{rest}}/c^2 > 0$ (Eq.(32) in Sect.3.1). Under this condition, we could demonstrate the run-away avoidance by using the Röhrlich method [16] (Sect.3.2), and we could perform the numerical calculation for checking the difference between the proposed models. The results showed the dependence of $q(\chi)$, the correction in high-intensity fields being the essential difference between the models (Sect.3.3). This equation only requires that the field strength of the external field is smaller than the Schwinger limit. Of course, this equation holds also the invariance $(dx^\mu/d\tau)(dx_\mu/d\tau) = c^2$ which the QED-Sokolov equation cannot satisfy. In the results of the numerical simulation and analysis, the proposed Eq.(43/44) can include the dynamics of QED-Sokolov's equation (4-6) as long as we choose the relation $\Xi = \Theta = q(\chi)$.

We introduce the measure of an electron's mass $m(x)$ and an anisotropic electron's charge $\mathcal{E}(x) \in \mathbb{V}_M^4 \otimes \mathbb{V}_M^4 \otimes \mathbb{V}_M^4 \otimes \mathbb{V}_M^4$ following Ref.[10].

$$\boxed{dm(x)\frac{dw^\mu}{d\tau} = -d\mathcal{E}^{\mu\nu}{}_{\alpha\beta}(x) \mathfrak{F}_{\text{hom}}{}^{\alpha\beta} w_\nu} \quad (45)$$

The Radon-Nikodym derivative [18] should be $d\mathcal{E}^{\mu\nu}{}_{\alpha\beta}/dm = e_{\text{High Field}}/m_{\text{High Field}} \times \mathcal{K}^{\mu\nu}{}_{\alpha\beta}$, depending on the invariance $\Xi$. The anisotropy of charge $\mathcal{E}(x)$ comes from the polarization of QED vacuum. This effect is a unique dynamics which the QED-Sokolov's equations (3-5) does not include. This Radon-Nikodym derivative $d\mathcal{E}/dm$ is the "QED vacuum filter of fields" between the dressed and undressed fields.

In Heisenberg-Euler vacuum [11,12] (Sect.3.1), we can find the limit of the photon energy(same as Ref.[9]). From the limit $1 - 2\eta \mathbf{E}_{\text{ex}}^2|_{\text{rest}}/c^2 > 0$, the field strength should satisfy $|\mathbf{E}_{\text{ex}}| < E_{\text{Schwinger}}$ with the definition of the Schwinger limit $E_{\text{Schwinger}} = m_0^2 c^3/e\hbar$. On the other hand, the limit of a photon energy comes from the expansion of the Heisenberg-Euler's action integral [9]:

$$\int d^4 x L_{\text{Heisenberg-Euler}}^{\text{1st order}}(\langle \mathfrak{F} | \mathfrak{F} \rangle, \langle \mathfrak{F} | *\mathfrak{F} \rangle)$$
$$= -\int d^4 x \frac{1}{4\mu_0} \langle \mathfrak{F} | \mathfrak{F} \rangle + \int d^4 x \frac{\alpha^2 \hbar^3 \varepsilon_0^2}{360 m_0^4 c} \left[ 4\langle \mathfrak{F} | \mathfrak{F} \rangle^2 + 7\langle \mathfrak{F} | *\mathfrak{F} \rangle^2 \right] + O\left( \left( \frac{g_{\mu\nu} \hbar k^\mu \hbar k^\nu}{m_0^2 c^2} \right)^3 \right)$$

(46)

For this expansion, $2g_{\mu\nu} \hbar k_{\text{laser}}{}^\mu \hbar k_{\text{radiation}}{}^\nu < 4 \times \hbar\omega_{\text{laser}} \times \hbar\omega_{\text{radiation}}/c^2 < m_0^2 c^2 = (0.5\,\text{MeV}/c)^2$. In the numerical simulation of the head-on collision, we used the laser wavelength of $0.82\,\mu\text{m}$ equivalent to 1.5 eV , therefore, the maximum radiated photon energy is $\hbar\omega_{\text{radiation}} < O(10\,\text{GeV})$ [9]. Those are the phenomenological limits of the Maxwell's equation in QED vacuum, in the proposed model. Exceeding this limit is equivalent to breaking QED vacuum with electron-positron pair creation by energetic photons.

Finally, we discuss the experiments for checking radiation reaction models. In this paper, we used the plausible relation $\Xi = q(\chi)$ which converges the dynamics to the Sokolov's equation [14]. Under $\tau_0 \, d\Xi/d\tau \ll 1$ for a simplification, we can observe the radiated field from an electron

$$F_{\text{Mod-LAD}}{}^{\mu\nu} \sim \Xi \times F_{\text{LAD}}{}^{\mu\nu}. \tag{47}$$

$\Xi$ can round the QED effects into the classical dynamics via Eqs.(8-9) by the present proposal. We may find the relation $F_{\text{Mod-LAD}}{}^{\mu\nu} \sim [q(\chi) + \delta q] \times F_{\text{LAD}}{}^{\mu\nu}$, where $\delta q$ denotes the alteration from the synchronization. In this case, we shall go back to the equation

$$\begin{aligned}\frac{dw^\mu}{d\tau} &= -\frac{e_{\text{High Field}}}{m_{\text{High Field}}} \frac{(1-\eta f_0)\mathfrak{F}_{\text{hom}}{}^{\mu\nu}w_\nu + \eta g_0(*\mathfrak{F}_{\text{hom}})^{\mu\nu}w_\nu}{(1-\eta f_0)^2 + (\eta g_0)^2} \\ &= -\frac{e}{m_0}\frac{\Xi}{\Theta} \frac{(1-\eta f_0)\mathfrak{F}_{\text{hom}}{}^{\mu\nu}w_\nu + \eta g_0(*\mathfrak{F}_{\text{hom}})^{\mu\nu}w_\nu}{(1-\eta f_0)^2 + (\eta g_0)^2}\end{aligned}, \tag{48}$$

before putting the relation $\Xi = \Theta$. The relation Eq.(47) is replaced by $F_{\text{Mod-LAD}}{}^{\mu\nu} \sim \Xi/\Theta \times F_{\text{LAD}}{}^{\mu\nu} = [q(\chi) + \delta q] \times F_{\text{LAD}}{}^{\mu\nu}$. Since QED-based synchrotron radiation is the general electromagnetic interaction between an electron and the external fields we can assume $\Xi = q(\chi)$ from the main discussion in this paper. Hence, $\Theta = \Xi/[q(\chi) + \delta q] \sim 1 - \delta q/q(\chi)$ is estimated. For example, one of the candidates of this effect is non-electromagnetic interactions like the Poincaré stress which is introduced by the inner structure of an electron for the stabilization of its electromagnetic mass [19], so we can also extend this model to unknown non-electromagnetic interactions. Since $d\mathcal{E}^{\mu\nu}{}_{\alpha\beta}/dm = e_{\text{High Field}}/m_{\text{High Field}} \times \mathcal{K}^{\mu\nu}{}_{\alpha\beta} = e/m_0 \times \Xi/\Theta \times \mathcal{K}^{\mu\nu}{}_{\alpha\beta}$ represents a QED coupling correction between an electron and radiation in high-intensity fields, the investigation of $\Xi$ and $\Theta$ will become more important for radiation reaction acting on an electron in ultrahigh-intensity fields.

**Acknowledgements**

This work is supported by Extreme Light Infrastructure – Nuclear Physics (ELI-NP) – Phase I, a project co-financed by the Romanian Government and the European Union through the European Regional Development Fund, and also partly supported under the auspices of the Japanese Ministry of Education, Culture, Sports, Science and Technology (MEXT) project on "Promotion of relativistic nuclear physics with ultra-intense laser."

**Appendix 1: Detail of the derivation of equation of motion Eq.(25)**

Here we derive the undressed field $\tilde{\mathfrak{F}} \in \mathbb{V}_M^4 \otimes \mathbb{V}_M^4$ from Eq.(23).

$$\tilde{\mathfrak{F}}^{\mu\nu} = \frac{(1-\eta f)\mathfrak{F}_{\text{hom}}{}^{\mu\nu} + \eta g (*\mathfrak{F}_{\text{hom}})^{\mu\nu}}{(1-\eta f)^2 + (\eta g)^2} \qquad (A\text{-}1.1)$$

The definition of the invariant function $f$ and $g$ are Eq.(20) and Eq.(21). We can expect the form of

$$\tilde{\mathfrak{F}}^{\mu\nu} = \mathfrak{F}_{\text{hom}}{}^{\mu\nu} + (\delta \times \theta)\mathfrak{F}_{\text{hom}}{}^{\mu\nu} + (\delta \times \theta_*)(*\mathfrak{F}_{\text{hom}})^{\mu\nu} . \qquad (A\text{-}1.2)$$

We assume that the parameter $\delta$ satisfies the relation $|\delta| \ll 1$. The functions $\theta$ and $\theta_*$ depend on $\tilde{\mathfrak{F}}$. This relation leads to the expansion of the invariant functions:

$$f(\langle \tilde{\mathfrak{F}} | \tilde{\mathfrak{F}} \rangle, \langle \tilde{\mathfrak{F}} | *\tilde{\mathfrak{F}} \rangle) = f(\langle \mathfrak{F}_{\text{hom}} | \mathfrak{F}_{\text{hom}} \rangle, \langle \mathfrak{F}_{\text{hom}} | *\mathfrak{F}_{\text{hom}} \rangle) + \delta \times \Theta_f \qquad (A\text{-}1.3)$$

$$g(\langle \tilde{\mathfrak{F}} | \tilde{\mathfrak{F}} \rangle, \langle \tilde{\mathfrak{F}} | *\tilde{\mathfrak{F}} \rangle) = g(\langle \mathfrak{F}_{\text{hom}} | \mathfrak{F}_{\text{hom}} \rangle, \langle \mathfrak{F}_{\text{hom}} | *\mathfrak{F}_{\text{hom}} \rangle) + \delta \times \Theta_g \qquad (A\text{-}1.4)$$

For instance, we introduce $f_0 = f(\langle \mathfrak{F}_{\text{hom}} | \mathfrak{F}_{\text{hom}} \rangle, \langle \mathfrak{F}_{\text{hom}} | *\mathfrak{F}_{\text{hom}} \rangle)$ and $g_0 = g(\langle \mathfrak{F}_{\text{hom}} | \mathfrak{F}_{\text{hom}} \rangle, \langle \mathfrak{F}_{\text{hom}} | *\mathfrak{F}_{\text{hom}} \rangle)$. By using these equations, Eq.(A-1.1) becomes

$$\tilde{\mathfrak{F}}^{\mu\nu} = \frac{(1-\eta f_0 - \eta\delta \times \Theta_f)\mathfrak{F}_{\text{hom}}{}^{\mu\nu} + (\eta g_0 + \eta\delta \times \Theta_g)(*\mathfrak{F}_{\text{hom}})^{\mu\nu}}{(1-\eta f_0)^2 + (\eta g_0)^2}$$

$$\times \sum_{n=0}^{\infty} \eta^n \delta^n \left[ \frac{2\Theta_f - 2\eta(f_0\Theta_f + g_0\Theta_g) - \eta\delta(\Theta_f^2 + \Theta_g^2)}{(1-\eta f_0)^2 + (\eta g_0)^2} \right]^n$$

$$= \frac{(1-\eta f_0)\mathfrak{F}_{\text{hom}} + \eta g_0 (*\mathfrak{F}_{\text{hom}})^{\mu\nu} + O(\eta\delta)}{(1-\eta f_0)^2 + (\eta g_0)^2} \qquad . \qquad (A\text{-}1.5)$$

By neglecting the terms of $O(\eta\delta)$,[XVI]

$$\boxed{\tilde{\mathfrak{F}}^{\mu\nu} = \frac{(1-\eta f_0)\mathfrak{F}_{\text{hom}} + \eta g_0 (*\mathfrak{F}_{\text{hom}})^{\mu\nu}}{(1-\eta f_0)^2 + (\eta g_0)^2}} , \qquad (A.1\text{-}6)$$

with the definition as follows:

$$\mathfrak{F}_{\text{hom}} = F_{\text{ex}} + F_{\text{Mod-LAD}} , \qquad (A.1\text{-}7)$$

$$F_{\text{Mod-LAD}}{}^{\mu\nu} = -\frac{m_0 \tau_0}{ec^2} \left[ \frac{d^2(\Xi w^\mu)}{d\tau^2} w^\nu - \frac{d^2(\Xi w^\nu)}{d\tau^2} w^\mu \right] . \qquad (A.1\text{-}8)$$

By substituting $\tilde{\mathfrak{F}}$ into the equation of motion

$$m_{\text{High Field}}(\tau) \frac{dw^\mu}{d\tau} = -e_{\text{High Field}}(\tau) \tilde{\mathfrak{F}}^{\mu\nu} w_\nu \qquad (A\text{-}1.9)$$

---

[XVI] This order cut-off is important for the stability of the new equation. See Sect. 3.1 and 3.2.

becomes

$$\frac{dw^\mu}{d\tau} = -\frac{e_{\text{High Field}}(\tau)}{m_{\text{High Field}}(\tau)} \frac{(1-\eta f_0)\mathfrak{F}_{\text{hom}}^{\mu\nu} w_\nu + \eta g_0 (*\mathfrak{F}_{\text{hom}})^{\mu\nu} w_\nu}{(1-\eta f_0)^2 + (\eta g_0)^2} . \quad \text{(A-1.10)}$$

In the case of the low intensity limit, the charge to mass ratio should be $e_{\text{High Field}}/m_{\text{High Field}} = e/m_0 = 1.75 \times 10^{11} [\text{C/kg}]$. How is it in the case of high-intensity fields? By transforming Eq.(A-1.10),

$$m_0 \frac{dw^\mu}{d\tau} = \frac{\Xi}{\Theta} \frac{(1-\eta f_0) \times [f_{\text{ex}}^\mu + \Xi \times f_{\text{LAD}}^\mu] + \eta g_0 (*f_{\text{ex}})^\mu}{(1-\eta f_0)^2 + (\eta g_0)^2 - 2\tau_0(1-\eta f_0)\frac{\Xi}{\Theta}\frac{d\Xi}{d\tau}} . \quad \text{(A-1.11)}$$

Following Sect.2.1, we choose the relation $\eta f_0$, $\eta g_0$ and $\Xi/\Theta \times \tau_0 d\Xi/d\tau$ is enough smaller than unity, $\Xi$ and $\Theta$ are the same order of the magnitude. By this choice we can reduce this equation like

$$m_0 \frac{dw^\mu}{d\tau} \approx \frac{\Xi}{\Theta} \times f_{\text{ex}}^\mu + \frac{\Xi^2}{\Theta} \times f_{\text{LAD}}^\mu = \frac{\Xi}{\Theta} \times f_{\text{ex}}^\mu + \frac{\Xi^2}{\Theta} \times m_0\tau_0 \frac{d^2 w^\mu}{d\tau^2} + \frac{\Xi^2}{\Theta} \times \frac{m_0\tau_0}{c^2} g_{\alpha\beta} \frac{dw^\alpha}{d\tau} \frac{dw^\beta}{d\tau} w^\mu$$

$$\text{(A-1.12)}$$

Therefore, the Larmor's radiation formula obtain the correction factor of $\Xi^2/\Theta$,

$$\left.\frac{dW}{dt}\right|_{\text{High Field}} = -\frac{\Xi^2}{\Theta} \times m_0 \tau_0 g_{\alpha\beta} \frac{dw^\alpha}{d\tau} \frac{dw^\beta}{d\tau} . \quad \text{(A-1.13)}$$

Following the QED-based synchrotron radiation formula,

$$\left.\frac{dW}{dt}\right|_{\text{High Field}} = q(\chi) \times \left.\frac{dW}{dt}\right|_{\text{Classical}} , \quad \text{(A-1.14)}$$

$$q(\chi) = \frac{9\sqrt{3}}{8\pi} \int_0^{\chi^{-1}} dr\, r \left[\int_{r_\chi}^\infty dr'\, K_{5/3}(r') + \chi^2 r r_\chi K_{2/3}(r_\chi)\right] , \quad \text{(A-1.15)}$$

we can find the candidate $\Xi^2/\Theta = q(\chi)$. In Eq.(A-1.12), $\Xi/\Theta \times f_{\text{ex}}$ should be $f_{\text{ex}}$ in QED-Sokolov model, $\Theta = \Xi$. Therefore, we can propose the relation $\Theta = \Xi \Rightarrow e_{\text{High Field}}/m_{\text{High Field}} = e/m_0$ and the equation of a radiating electron's motion,

$$\boxed{\frac{dw^\mu}{d\tau} = -\frac{e}{m_0} \frac{(1-\eta f_0)\mathfrak{F}_{\text{hom}}^{\mu\nu} w_\nu + \eta g_0 (*\mathfrak{F}_{\text{hom}})^{\mu\nu} w_\nu}{(1-\eta f_0)^2 + (\eta g_0)^2}} \quad \text{(A-1.16)}$$

Conversely, the choices of $\Theta = \Xi$ and $\Xi = q(\chi)$ satisfy the relations, $\eta f_0$, $\eta g_0 \ll 1$ and $\Xi/\Theta \times |\tau_0 d\Xi/d\tau| \ll 1$.

**Appendix 2: Errata of K. Seto, PTEP 2015 [10]**

Strictly speaking, Eq.(1) should be

$$\frac{dw^\mu}{d\tau} \stackrel{\Xi \equiv 1}{=} -\frac{e}{m_0} \frac{[(1-\eta f_0)\mathfrak{F}^{\mu\nu} + \eta g_0 (*\mathfrak{F})^{\mu\nu}]w_\nu}{(1-\eta f_0)^2 + (\eta g_0)^2} \quad , \tag{A.2-1}$$

for connecting from Eq.(44) or Eq. (A.1-19). Here, $\mathfrak{F} = \mathfrak{F}_{\text{hom}}|_{\Xi \equiv 1} = F_{\text{ex}} + F_{\text{LAD}} \in \mathbb{V}_M^4 \otimes \mathbb{V}_M^4$. Normally, $|1-\eta f_0| \gg |\eta g_0|$ is satisfied, the Eq.(A.2-1) is transformed as follows under this condition:

$$\frac{dw^\mu}{d\tau} \stackrel{\Xi \equiv 1}{=} -\frac{e}{m_0(1-\eta f_0)} \frac{\mathfrak{F}^{\mu\nu} + \frac{\eta g_0}{1-\eta f_0}(*\mathfrak{F})^{\mu\nu}}{1 + \frac{(\eta g_0)^2}{(1-\eta f_0)^2}} w_\nu = -\frac{e}{m_0(1-\eta f_0)}[\mathfrak{F}^{\mu\nu} + \eta g_0(*\mathfrak{F})^{\mu\nu}]w_\nu + \cdots$$

(A.2-2)

This is the Eq.(1) derived in Ref.[10]. However in the strict order expansion, it should be

$$\frac{dw^\mu}{d\tau} \stackrel{\Xi \equiv 1}{=} -\frac{e}{m_0(1-\eta f_0)} \mathfrak{F}^{\mu\nu} w_\nu + \mathrm{O}\left(\frac{\eta g_0}{1-\eta f_0}\right). \tag{A.2-3}$$

In this form, the analysis of run-away avoidance in Ref.[10] becomes easier[XVII], the numerical calculation almost agreed since $|\mathfrak{F}^{\mu\nu}| \gg |\eta g_0 (*\mathfrak{F})^{\mu\nu}|$ is satisfied. We suggested the anisotropy of the QED vacuum. We can confirm the anisotropic field $*\mathfrak{F}$ in the form of Eq.(A.2-1), but it does not exist in (A.2-3). The higher orders of $\eta g_0/(1-\eta f_0)$ describe the degree of the anisotropy of QED vacuum.

---

[XVII] For considering Eq.(A.2-2), We only put the relation $g_0 = 0$ at Ch.3 in Ref.[10] and any anisotropy is vanished.

**Appendix 3: the detail of Eq.(35)**

From Eq(34),

$$m_0\tau_0 g_{\mu\nu}\frac{dw^\mu}{d\tau}\frac{dw^\nu}{d\tau} = \frac{\tau_0}{m_0}\frac{\frac{e^2c^2}{2\eta}(1-\eta f_0)^2 \eta f_0 + \frac{2e^2c^2}{7\eta}(1-\eta f_0)(\eta g_0)^2}{|(1-\eta f_0)^2+(\eta g_0)^2|^2}$$
$$+\frac{\tau_0}{m_0}\frac{g_{\mu\nu}[(1-\eta f_0)f_{ex}{}^\mu + \eta g_0 * f_{ex}{}^\mu][(1-\eta f_0)f_{ex}{}^\nu + \eta g_0 * f_{ex}{}^\nu]}{|(1-\eta f_0)^2+(\eta g_0)^2|^2}.$$

(A.3-1)

We consider this equation under the condition of Eq.(32);

$$1-\eta f_0 \overset{\text{physical requirements}}{>} 0.$$

(A.3-2)

This condition supports the relation $(1-\eta f_0)^2 + (\eta g_0)^2 > 0$. Considering the above, we can proceed to consider the absolute value of Eq.(A.3-1).

$$\left|m_0\tau_0 g_{\mu\nu}\frac{dw^\mu}{d\tau}\frac{dw^\nu}{d\tau}\right| = \frac{\tau_0}{m_0}\frac{\frac{e^2c^2}{2\eta}(1-\eta f_0)^2|\eta f_0| + \frac{2e^2c^2}{7\eta}(1-\eta f_0)(\eta g_0)^2}{|(1-\eta f_0)^2+(\eta g_0)^2|^2}$$
$$+\frac{\tau_0}{m_0}\frac{|g_{\mu\nu}[(1-\eta f_0)f_{ex}{}^\mu + \eta g_0(*f_{ex})^\mu][(1-\eta f_0)f_{ex}{}^\nu + \eta g_0(*f_{ex})^\nu]|}{|(1-\eta f_0)^2+(\eta g_0)^2|^2}$$
$$\leq \frac{\tau_0}{m_0}\frac{\frac{e^2c^2}{2\eta}(1-\eta f_0)^2|\eta f_0| + \frac{2e^2c^2}{7\eta}(1-\eta f_0)(\eta g_0)^2}{(1-\eta f_0)^4}$$
$$+\frac{\tau_0}{m_0}\frac{|g_{\mu\nu}[(1-\eta f_0)f_{ex}{}^\mu + \eta g_0(*f_{ex})^\mu][(1-\eta f_0)f_{ex}{}^\nu + \eta g_0(*f_{ex})^\nu]|}{(1-\eta f_0)^4}$$
$$\leq \frac{\tau_0}{m_0}\frac{\frac{e^2c^2}{2\eta}|\eta f_0|}{(1-\eta f_0)^2} + \frac{\tau_0}{m_0}\frac{\frac{2e^2c^2}{7\eta}(\eta g_0)^2}{(1-\eta f_0)^3}$$
$$+\frac{\tau_0}{m_0}\frac{|g_{\mu\nu}f_{ex}{}^\mu f_{ex}{}^\nu|}{(1-\eta f_0)^2} + \frac{2\tau_0}{m_0}\frac{|\eta g_0|}{1-\eta f_0}\times\frac{|g_{\mu\nu}f_{ex}{}^\mu(*f_{ex})^\nu|}{(1-\eta f_0)^2}$$
$$+\frac{\tau_0}{m_0}\frac{\eta g_0}{1-\eta f_0}\frac{|g_{\mu\nu}f_{ex}{}^\mu(*f_{ex})^\nu|}{(1-\eta f_0)^2}$$

(A.3-3)

Finally, $O(|g_0|) < O(|f_0|)$ is satisfied in the case of run-away, we can obtain the relation of Eq.(35).

$$\left| m_0 \tau_0 g_{\mu\nu} \frac{dw^\mu}{d\tau} \frac{dw^\nu}{d\tau} \right|$$

$$\stackrel{\text{run-away}}{<} \frac{\tau_0}{m_0} \frac{e^2 c^2}{2\eta} \frac{|\eta f_0|}{(1-\eta f_0)^2} + \frac{\tau_0}{m_0} \frac{2e^2 c^2}{7\eta} \frac{1}{1-\eta f_0} \frac{(\eta f_0)^2}{(1-\eta f_0)^2}$$

$$+ \frac{\tau_0}{m_0} \frac{|g_{\mu\nu} f_{\text{ex}}{}^\mu f_{\text{ex}}{}^\nu|}{(1-\eta f_0)^2} + \frac{2\tau_0}{m_0} \frac{|\eta f_0|}{1-\eta f_0} \times \frac{|g_{\mu\nu} f_{\text{ex}}{}^\mu (*f_{\text{ex}})^\nu|}{(1-\eta f_0)^2}$$

$$+ \frac{\tau_0}{m_0} \frac{\eta g_0}{(1-\eta f_0)^2} \frac{|g_{\mu\nu} f_{\text{ex}}{}^\mu (*f_{\text{ex}})^\nu|}{1-\eta f_0}$$

(A.3-4)